\documentclass[10.5pt,compsoc]{tst}
\usepackage{graphicx}
\usepackage{footmisc}
\usepackage{subfigure}
\usepackage{url}
\usepackage{multirow}
\usepackage[noadjust]{cite}
\usepackage{amsmath,amsthm}
\usepackage{amssymb,amsfonts}
\usepackage{booktabs}
\usepackage{color}
\usepackage{ccaption}
\usepackage{booktabs}
\usepackage{float}
\usepackage{fancyhdr}
\usepackage{caption}
\usepackage{xcolor,stfloats}
\usepackage{comment}
\graphicspath{{figures/}}
\usepackage{cuted}  
\usepackage{captionhack}
\usepackage{epstopdf}

\usepackage{algorithm}
\usepackage{algorithmic}

\headevenname{\normalsize{\textbf{\emph{Tsinghua Science and Technology, xxxxx}} 20xx, 2x(x): xxx--xxx}}%
\headoddname{{\sf First author et al.:}\quad {\textbf{\emph{Click and Type Your Title Here ...}}}}%

\newtheoremstyle{mystyle}{0pt}{0pt}{\normalfont}{1em}{\bf}{}{1em}{}
\theoremstyle{mystyle}
\renewcommand\figurename{Fig.~}

\newtheorem{definition}{\textbf{Definition}}

\newcommand{\nop}[1]{}

\makeatletter
\renewcommand{\@biblabel}[1]{[#1]\hfill}
\makeatother

\begin{document}

\clearpage

\hyphenpenalty=50000

\makeatletter
\newcommand\mysmall{\@setfontsize\mysmall{7}{9.5}}

\newenvironment{tablehere}
  {\def\@captype{table}}
  {}
\newenvironment{figurehere}
  {\def\@captype{figure}}
  {}

\thispagestyle{plain}%
\thispagestyle{empty}%

\let\temp\footnote
\renewcommand \footnote[1]{\temp{\normalsize #1}}
{}
\vspace*{-40pt}
\noindent{\normalsize\textbf{\scalebox{0.885}[1.0]{\makebox[5.9cm][s]
{TSINGHUA\, SCIENCE\, AND\, TECHNOLOGY}}}}

\vskip .2mm
{\normalsize
\textbf{
\hspace{-5mm}
\scalebox{1}[1.0]{\makebox[5.6cm][s]{%
I\hfill S\hfill S\hfill N\hfill{\color{white}%
l\hfill l\hfill}1\hfill0\hfill0\hfill7\hfill-\hfill0\hfill2\hfill1\hfill4
\hfill \color{white}{\quad 0\hfill ?\hfill /\hfill ?\hfill ?\quad p\hfill p\hfill  ?\hfill ?\hfill ?\hfill --\hfill ?\hfill ?\hfill ?}\hfill}}}}

\vskip .2mm
{\normalsize
\textbf{
\hspace{-5mm}
\scalebox{1}[1.0]{\makebox[5.6cm][s]{%
DOI:~\hfill~\hfill1\hfill0\hfill.\hfill2\hfill6\hfill5\hfill9\hfill9\hfill/\hfill T\hfill S\hfill T\hfill.\hfill2\hfill0\hfill x\hfill x\hfill.\hfill9\hfill0\hfill1\hfill0\hfill x\hfill x\hfill x}}}}

\vskip .2mm\noindent
{\normalsize\textbf{\scalebox{1}[1.0]{\makebox[5.6cm][s]{%
\color{black}{V\hfill o\hfill l\hfill u\hfill m\hfill%
e\hspace{0.356em}xx,\hspace{0.356em}N\hfill u\hfill%
m\hfill b\hfill e\hfill r\hspace{0.356em}x,\hspace{0.356em}%
x\hfill x\hfill x\hfill x\hfill x\hfill%
x\hfill x\hfill \hspace{0.356em}2\hfill0\hfill x\hfill x}}}}}\\

\begin{strip}
{\center
{\LARGE\textbf{Betweenness Approximation for Edge Computing with Hypergraph Neural Network}}
\vskip 9mm}

{\center {\sf \large
Yaguan Guo, Wenxin Xie, Qingren Wang$^*$, Dengcheng Yan, and Yiwen Zhang
}
\vskip 5mm}

\centering{
\begin{tabular}{p{160mm}}

{\normalsize
\linespread{1.6667} %
\noindent
\bf{Abstract:} {\sf
Edge computing is highly demanded to achieve their full potentials Internet of Things (IoT), since various IoT systems have been generating big data facilitating modern latency-sensitive applications. As a basic problem, network dismantling tries to find an optimal set of nodes of which will maximize the connectivity degradation in a network. However, current approaches mainly focus on simple networks modeling only pairwise interactions between two nodes, while higher order groupwise interactions among arbitrary number of nodes are ubiquitous in real world which can be better modeled as hypernetwork. The structural difference between simple network and hypernetwork restricts the direct application of simple network dismantling methods to hypernetwork. Even though some hypernetwork centrality measures such as betweenness can be used for hypernetwork dismantling, they face the problem of balancing effectiveness and efficiency. Therefore, we propose a betweenness approximation-based hypernetwork dismantling method with hypergraph neural network, namely HND. HND trains a transferable hypergraph neural network-based regression model on plenty of generated small-scale synthetic hypernetwork in a supervised way, and utilizes the well-trained model to approximate nodes' betweenness. Extensive experiments on five real hypernetworks demonstrate the effectiveness and efficiency of HND comparing with various baselines.}
\vskip 4mm
\noindent
{\bf Key words:} {\sf hypernetwork dismantling; graph neural network; betweenness approximation; edge computing}}

\end{tabular}
}
\vskip 6mm

\vskip -3mm
\small\end{strip}

\thispagestyle{plain}%
\thispagestyle{empty}%
\makeatother
\pagestyle{tstheadings}

\begin{figure}[b]
\vskip -6mm
\begin{tabular}{p{44mm}}
\toprule\\
\end{tabular}
\vskip -4.5mm
\noindent
\setlength{\tabcolsep}{1pt}
\begin{tabular}{p{1.5mm}p{79.5mm}}
$\bullet$& Yaguang Guo is with the School of Management, Hefei University of Technology, Hefei and 230009, China. E-mail: 2005800134@hfut.edu.cn\\
$\bullet$& Wenxin Xie, Qingren Wang, Dengcheng Yan and Yiwen Zhang are with the School of Computer Science and Technology, Anhui University, Hefei and 230601, China. E-mail: xiewxahu@foxmail.com, \{wqr, yanzhou, zhangyiwen\}@ahu.edu.cn\\
$\sf{*}$&
To whom correspondence should be addressed. \\
          &          Manuscript received: year-month-day;
          accepted: 2023-Sep-27

\end{tabular}
\end{figure}\large

\vspace{3.5mm}
\section{Introduction}
\label{s:introduction}
\noindent
Internet of Things (IoT) has significantly been changing the way we live \cite{qi_papoi, liu_alstm}, from a variety of aspects including entertainment, agriculture, manufacturing, etc. In this situation, the recently emerged new computing paradigms, e.g., Edge Computing has provided a beneficial complement to the traditional cloud-based systems, as edge can provide partial computing resources closer to the user or device side. Intuitively, the Internet of Things is a heterogeneous service network, whereas network has the ability of depicting entities and their relations \cite{sttl}, so that it is widely applied to model various IoT application systems, which forms one of the major sources of data facilitating modern latency-sensitive applications at the network edge like artificial intelligence, industrial automation, smart transportation. Hence, edge computing is highly demanded to achieve their full potentials IoT.

However, as a basic problem in network science, network dismantling (ND) \cite{nd} aims to find a set of nodes whose removal will greatly destory the connectivity of network. Due to the network connectivity is highly related to spread efficiency of information or infectious disease, ND has a widely application in corresponding fields \cite{epidemic, rumour}. Many researchers are attracted and multiple dismantling methods are proposed to solve this problem. However, current methods \cite{nd, gnd, finder} are mainly focus on the traditional simple network which only considers the pairwise relation between two nodes, while there are lots of higher-order groupwise relations among arbitrary number of nodes in real world and traditional simple network cannot model this kind of relation. Thus, the applicatin of these methods is limited when faces the groupwise relation.

\begin{figure}[htbp!]
	\centering
	\subfigtopskip=0pt
	\subfigure[{Simple network}]{
		\includegraphics[scale=0.22]{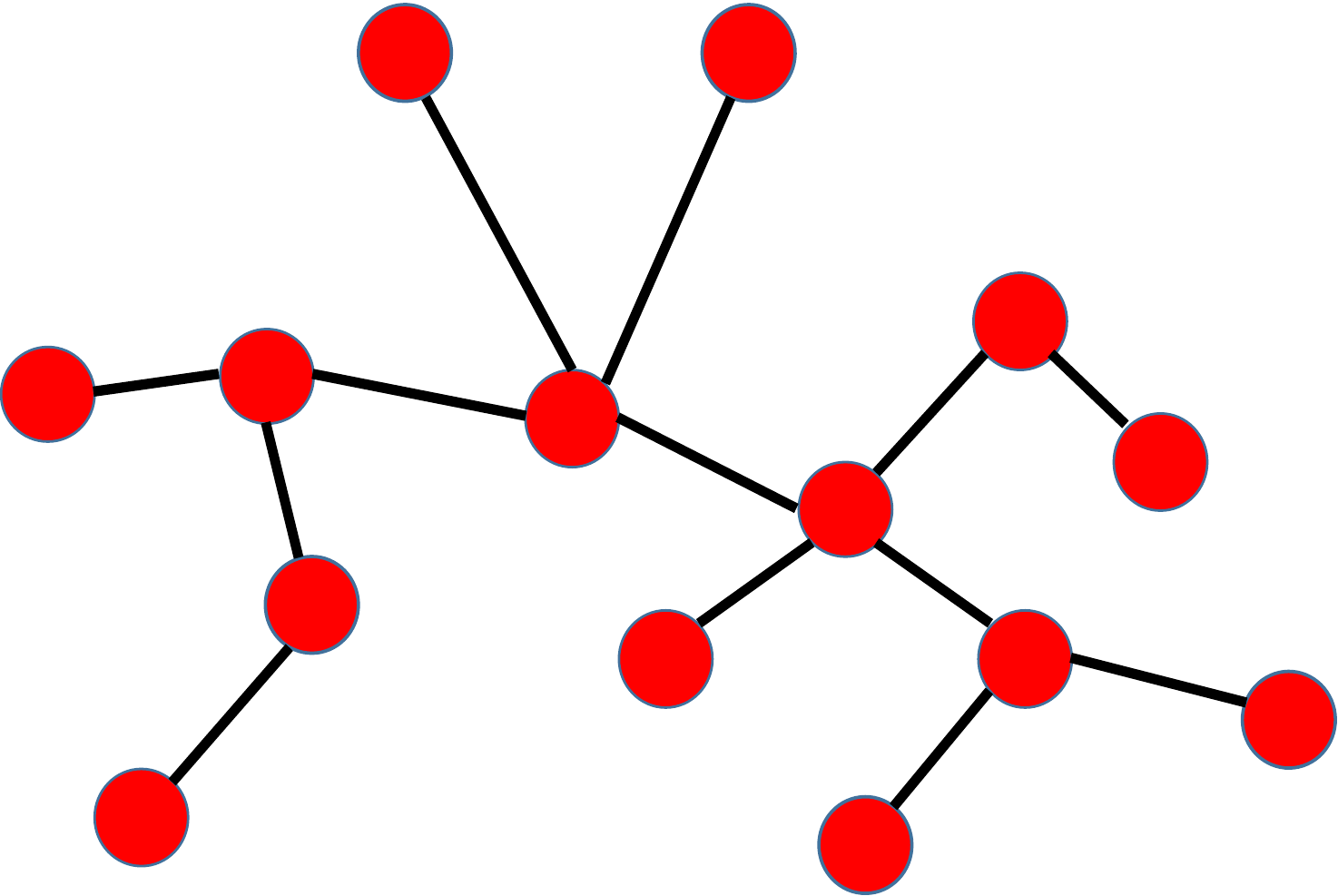}
		\label{subfig:simple network}
	}
	\subfigure[Hypernetwork]{
		\includegraphics[scale=0.22]{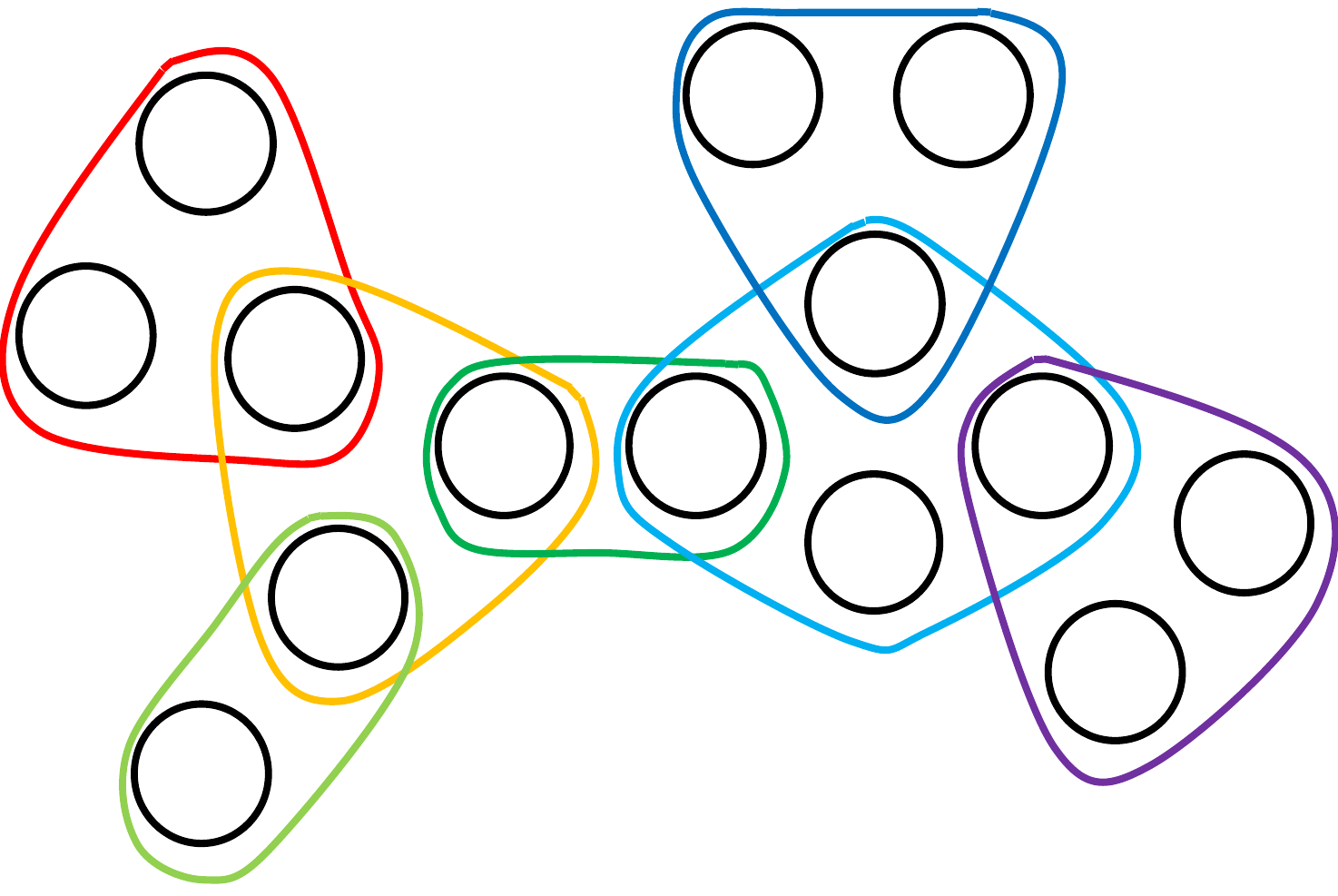}
		\label{subfig:hypernetwork}
	}
	\caption{Structural difference between hypernetwork and simple network}
	\label{fig:hypernetwork-and-simple-network}
\end{figure}

Fortunately, hypernetwork \cite{hypergraph} having unique advantages in modeling such groupwise relation is attracting many researchers. As shown in Fig.\ref{fig:hypernetwork-and-simple-network}, a hyperedge in hypernetwork can naturally express the higher-order relation among arbitrary number of nodes. Thus, utilizing hypernetwork to model real world systems is more suitable than traditional simple network when facing groupwise relations. However, even the higher-order relation modeling problem is solved by hypernetwork, the existing dismantling methods still have limitations when applying to hypernetwork. On the one hand, these methods cannot be directly applied due to the structural difference between hypernetwork and traditional simple network. On the other hand, although various centrality measures in hypernetwork can be used to dismantling just like in  traditional network, both of them face the problem of balancing effect and efficiency. For example, betweenness centrality is very suitable for dismantling a network but its calculation has a huge complexity. Degree centrality is easy to calculated but it performs not well on dismantling.

Recently, with the emergence of deep learning technology, researchers have attempted to utilize deep models to approximate complex centrality measures, such as betweenness \cite{bet-appro} and closeness \cite{clos-appro}. These measures can be approximated with less computational complexity in a limited error range. That is, the approximate centrality values are naturally utilized to various tasks so as to take both effect and efficiency into consideration. Therefore, this paper proposes a novel \textbf{H}yper\textbf{N}etwork \textbf{D}ismantling method based on deep learning technology, namely \emph{HND}. This method adopts the hypergraph neural network to approximate betweenness centrality in hypernetwork, and utilizes the approximate values to accomplish the hypernetwork dismantling task. Specifically, our proposed HND first generates many small scale synthetic hypernetworks and constructs betweenness ranking samples according to them. Then, a hypergraph neural network-based betweenness ranking model is built and the samples generated in previous step are used to trained this ranking model. Finally, the well-trained model is used to approximate the betweenness of all nodes in a given hypernetwork, and the hypernetwork is then dismantled according the approximate betweenness values.

Our main contributions are summarized as follows:
\begin{itemize}
	\item We design a betweenness approximation model based on hypergraph neural network. The model can be trained with lots of synthetic ranking samples and applied to real world hypernetworks. With the help of deep learning's powerful representation ability, the trained model can well approximate betweenness of real world hypernetwork with a lower computation complexity.
	\item We propose a novel hypernetwork dismantling method, called HND. HND utilizes the betweenness approximation model to calculate the approximate betweenness which is adopted to dismantle hypernetworks. Due to the approximation ability of our model, HND can achieve a performance close to betweenness with much lower computational complexity.
	\item We conduct extensive experiments on five real world hypernetworks. The results show that the performance of our proposed method outperforms the baselines. Moreover, the experimental results also confirm that the betweenness ranking model can approximate betweenness with less time consumption.
\end{itemize}

The rest of the paper is summarized as follows. Section \ref{sec:related work} introduces the related works about network dismantling and graph neural network. Section \ref{sec:preliminaries} gives some necessary definitions. In Section \ref{sec:method}, we introduce the proposed method in detail. Section \ref{sec:experiments} describes the experimental settings and demonstrates detailed analyses of the experimental results. In Section \ref{sec:conclusion}, we summarize the whole paper and give a potential direction of our future work.

\section{Related Work}
\label{sec:related work}

\subsection{Network Dismantling}
\label{sec:related work-network dismantling}
Network dismantling \cite{nd} aims to find an optimal set of nodes whose removal will greatly destroy the network connectivity. As a graph combinatorial optimization problem, it is NP-hard and its exact solution is hard to obtain. Thus, researchers attempt to find approximate solutions and propose various network dismantling methods. Generally, current network dismantling methods can be divided into three classes. The first class is based on centrality measures. In this kind of methods, nodes are selected greedily according to their centrality measures. However, these methods usually face the problem of balancing effectiveness and efficiency. Local centrality measures (e.g., degree centrality) are easy to calculate but cannot achieve a well dismantling performance, while global centrality measures (e.g., betweenness centrality, closeness centrality) perform well on dismantling but have a high computation complexity. In order to take both the effectiveness and efficiency into consideration, some centrality measures utilizing mesoscopic network structures are proposed. For example, the collective influence (CI) \cite{CI} proposed by Morone et al. can flexibly balance the globality and locality through  a tuning hyperparameter, so as to consider both the dismantling effectiveness and efficiency. The second class is heuristic methods. This kind of methods dismantle a network by multiple heuristic steps. For example, Braunstein et al. proposed a three step method MinSum \cite{nd} which dismantles a network by decycling, tree breaking and nodes reinsertion. Similar to MinSum, CoreHD \cite{corehd} and BPD \cite{bpd} are also adopt this framework but are different in details. Moreover, Ren et al. proposed the GND \cite{gnd} algorithm to consider the case of weighted nodes. The third class methods are based on deep learning. Researchers hope to achieve better dismantling effectiveness with the help of powerful ability of deep learning. Specifically, Fan et al. proposed FINDER \cite{finder} based on deep reinforcement learning. This method trains an agent to do dismantling exercises on lots of small scale synthetic networks, and finally applied to real large scale networks. Besides, Grassia et al. proposed the GDM \cite{gdm} method which trains an graph neural network ranking model by lots of ground-truth dismantling sequences and applied to real network dismantling. Recently, with the surge of hypernetwork, Yan et al. proposed a dismantling method suitable for hyernetworks based on deep reinforcement learning, called HITTER \cite{hitter}. Generally, current works mainly focus on traditional simple network but ignore the hypernetwork. Thus, in this paper we attempt to solve the hypernetwork dismantling problem via deep leaning technology.

\subsection{Graph Neural Network}
\label{sec:related work-graph neural network}
As a kind of network embedding method, graph neural network (GNN) can map nodes into low-rank vectors according to specific tasks. Based on the most basic neighbors information aggregation mechanism, various GNNs are proposed and applied to many fields. Among these GNNs, graph convolution network (GCN) \cite{GCN, liu_ietg} is usually seemed as a basic method. Through neighbor information aggregation, feature linear transformation and nonlinear activation, nodes' embedding can be obtained and applied to various down-stream tasks such as node classification, line prediction and so on. Other GNNs following this framework are proposed successively. For example, graph attention network (GAT) \cite{GAT} thought that different neighbors have different importance to target node in the step of neighbor aggregation, and it introduced the attention mechanism to make target node adaptively aggregation information. Hamilton et al. proposed inductive GraphSAGE \cite{graphsage} which can infrence the embeddings of nodes unseen in training stage. Moreover, due to its powerful expression ability, GNNs also have been applied to various fields such as recommendation systems \cite{ngcf, lightgcn, hsr}, user profile \cite{hgat, rup} and so on. However, the core neighbors information aggregation mechanism in GNN relies on the pairwise interections. Therefore, these GNNs cannnot be directly applied to hypernetworks due to the structural difference with traiditional simple network. In order to solve this problem, researchers proposed various methods to extend the traditional GNNs to hypernetworks. For example, Feng et al. proposed HGNN \cite{HGNN} to applied GCN to hypernetworks through transforming hypernetworks into simple networks according to clique expansion. Similarily, Yadati et al. \cite{HyperGCN} also transform hypernetworks into simple networks through breaking hyperedges into pairwise edges by specific rules, and directly applied GCN to it. Different from the above two methods, UniGNN \cite{UniGNN} proposed by Huang et al. gives the neighbor information aggregation mechanism suitable for hypernetworks. Through the aggregation paths of nodes to hyperedges and hyperedges to nodes, various traditional GNNs such as GCN, GAT and GIN \cite{gin} can be transfered to hypernetworks. In this paper, we adopt hypergrap neural network to accomplish betweenness approximation task. Although some works have attempted to approximate betweenness by GNNs, they are not suitable for hypernetworks. Therefore, a hypergraph neural network for hypernetwork betweenness approximation is necessary.

\section{Preliminaries}
\label{sec:preliminaries}
In this section, we briefly introduce some related concepts.

\begin{definition}[\textbf{Hypernetwork} \cite{hypergraph}]
	A hypernetwork is defined as $G=(V, E)$. The $V=\{v_1, v_2, \dots, v_N\}$ denotes node set in hypernetwork and $N=|V|$ is the number of nodes. The $E=\{e_1, e_2, \dots, e_M\}$ is the set of hyperedges and $M=|E|$ denotes the number of hyperedges. Each hyperedge $e$ is defined as $e \subseteq V$ and $e \ne \phi$.
\end{definition}
For each hyperedge $e$ in hypernetwork, the size of $e$ denotes the number of nodes contained in it. Obviously, due to the size of hyperedges is flexible, hypernetwork can model both pairwise and groupwise interactions. When the size of all hyperedges in a hypernetwork equals to 2, this hypernetwork becomes a traditional simple network. Therefore, traditional simple network can be seen as a specific form of hypernetwork.
\begin{definition}[\textbf{Incidence Matrix} \cite{2-section}]
	Incidence matrix of a hypernetwork is a matrix $\mathbf{H} \in \mathbb{R}^{N \times M}$. Each element of $\mathbf{H}$ is given as follows:
	\begin{equation}
		\label{equ:incidence matrix}
		\mathbf{H}_{v_i, e_j}=\left\{
		\begin{aligned}
			&1, \ \ \ \ \ \ \ \ v_i \in e_j \\
			&0, \ \ \ \ \ otherwise
		\end{aligned}
		\right.
	\end{equation}
\end{definition}
\begin{figure}[htbp!]
	\centering
	\subfigtopskip=0pt
	\subfigure[{IncidenceMatrix}]{
		\includegraphics[scale=0.22]{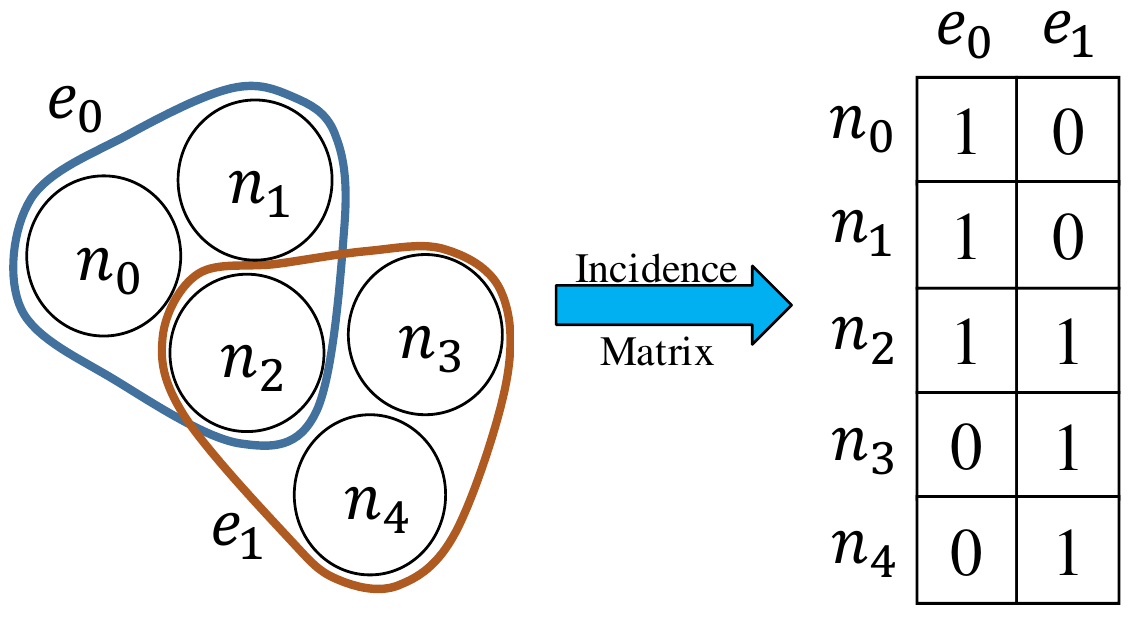}
		\label{subfig:incidence matrix}
	}
	\subfigure[Connectivity]{
		\includegraphics[scale=0.22]{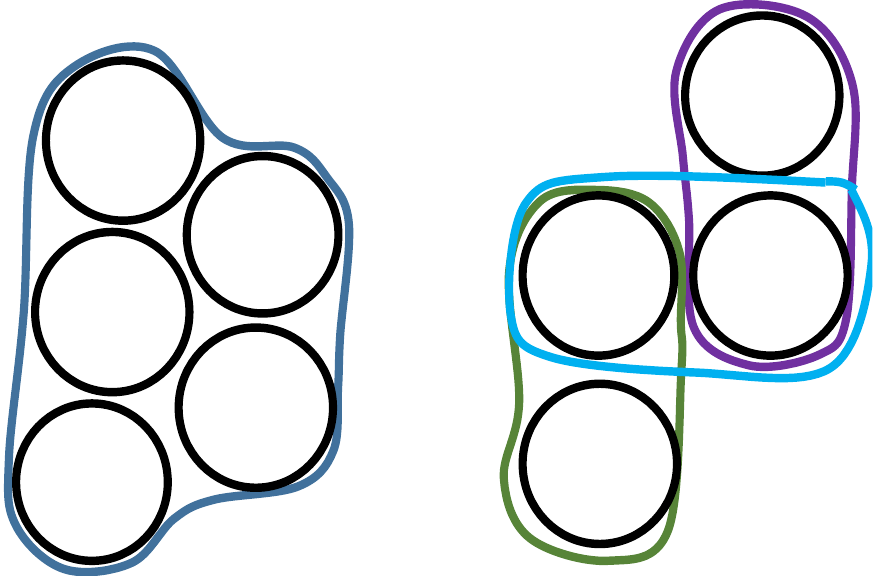}
		\label{subfig:connectivity}
	}
	\subfigure[2-section network]{
		\includegraphics[scale=0.24]{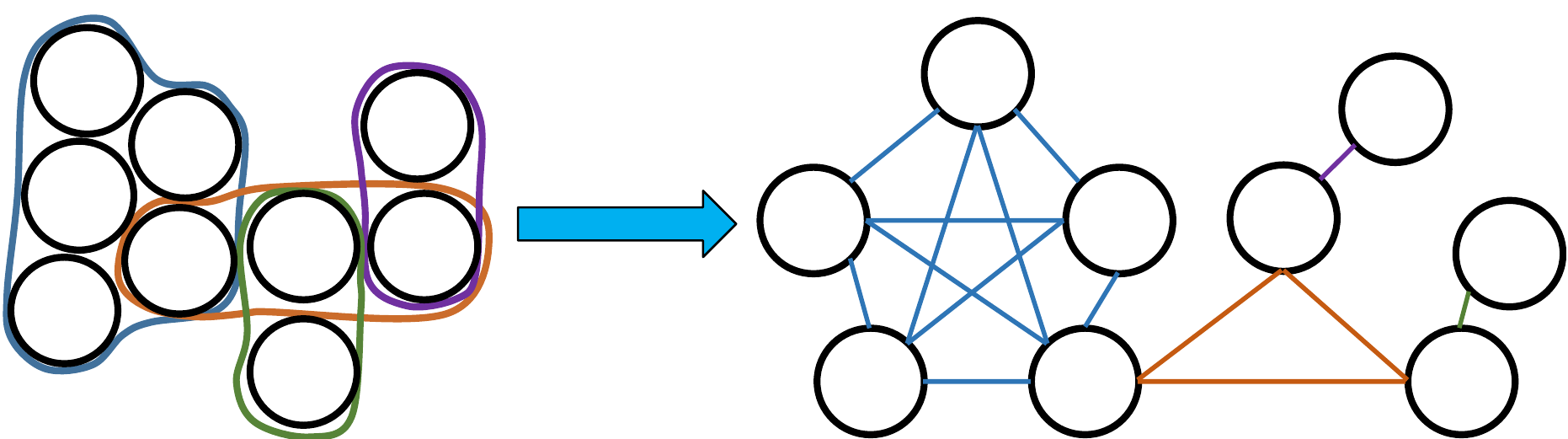}
		\label{subfig:2-section}
	}
	\caption{Hypernetwork and its corresponding definitions}
	\label{fig:hypernetwork and simple network}
\end{figure}
Similar to adjacent matrix in traditional simple network, incidence matrix can be used to express relations between nodes and hyperedges in hypernetwork. Fig. \ref {subfig:incidence matrix} shows an example hypernetwork and its incidence matrix.
\begin{definition}[\textbf{Hyperdegree} \cite{2-section} and \textbf{Degree} \cite{hypergraph}]
	For each node $v_i$ in hypernetwork $G$, the hyperdegree of $v_i$ is defined the number of hyperedges which contain node $v_i$:
	\begin{equation}
		\label{equ:hyperdegree}
		hdeg(v_i)=\sum_{e_j \in E}\mathbf{H}_{v_i, e_j}
	\end{equation}
	the degree denotes the number of nodes adjacent to it:
	\begin{equation}
		\label{equ:degree}
		deg(v_i)=\sum_{v_j \in V}(\mathbf{HH}^T)_{v_i, v_j} - hdeg(v_i)
	\end{equation}
\end{definition}
Due to the dismantling problem is highly related to network connectivity, so we define the connectivity of hypernetwork as follows:
\begin{definition}[\textbf{Hypernetwork connectivity}]
	The connected component contains the most number of hyperedges in hypernetwork $G$ is called the giant connected component (GCC). And the connectivity of $G$ is defined as the ratio of number of  nodes in GCC to the number of nodes in $G$:
	\begin{equation}
		\label{equ:connectivity}
		connectivity(G)=\frac {|V_{GCC}|} {|V_G|}
	\end{equation}
\end{definition}
Different from traditional simple network, our definition of hypernetwork connectivity is related the connections between hyperedges. According to Berge \cite{connectivity}, what a connected hypernetwork relies on is relations among hyperedges, not nodes. Moreover, it seems that selecting the connected component which contains the most number of nodes as the GCC is also reasonable. However, the connectivity defined in this way will be influenced by those huge hyperedges. As shown in Figure \ref{subfig:connectivity}, selecting the left connected component cannot reflect the property of hypernetwork connectivity, and it also violates the purpose of dismantling.
\begin{definition}[\textbf{2-section network} \cite{2-section}]
	2-section network is a traditional simple network transformed from a hypernetwork. For each hyperedge $e$ in a hypernetwork $G$, through linking each two nodes in $e$, the groupwise interaction can be transformed into multiple pairwise interactions, and the 2-section network is obtained when all hyperedges are transformed over.
\end{definition}
As shown in Fig. \ref{subfig:2-section}, each hypernetwork can transform into a simple network. In this way, various methods designed for simple network can be applied to hypernetwork.

\section{Proposed Method: HND}
\label{sec:method}
\subsection{Overall Framework}
\label{sec:method-overall framework}
In this section, we will introduce our proposed method in detail. Generally, our method can be divided into three steps.
	
\textbf{Step One}, to supply training samples for subsequent model, we adopt hypernetwork generator to generate lots of small scale synthetic hypernetworks, and calculate each node's betweenness value in them. According to the ground-truth betweenness values, training samples are constructed (as shown in Subsection \ref{sec:method-training samples generation}). 

\textbf{Step Two}, based on hypergraph neural network, we build a node betweenness approximation model in hypernetwork. This model can be applied to approximate node's betweenness in hypernetwork according to the network structure and predict nodes' approximation betweenness values (as shown in Subsection \ref{sec:method-node approximation model}).

\textbf{Step Three}, combining nodes ranking samples and approximation betweenness values, the pairwise ranking loss are built to optimize parameters in ranking model. After multiple iterations, the model can be used to approximate nodes' betweenness in real world hypernetworks and further applied to hypernetwork dismantling (as shown in Subsection \ref{sec:method-optimization}).

\subsection{Training Sample Generation}
\label{sec:method-training samples generation}
In this step, lots of node pair samples are generated according to their betweenness values. Thus, a synthetic hypernetwork generator is needed to generate small scale hypernetworks. There are two recently proposed synthetic hypernetwork generator, i.e., HyperPA \cite{hyperpa} and HyperFF \cite{hyperff}. The main difference between them are that HyperPA generates hypernetwork according to a predefined distribution from real world hypernetwork, while HyperFF adopts two hyper-parameters (i.e., burning probability $p$ and expanding probability $q$) to tune the density of synthetic hypernetworks. Comparing with HyperPA, HyperFF is more flexible and hypernetworks generated by it are more generalized. So HyperFF is adopted to accomplish this task.

Once the generator is chosen, lots of small scale synthetic hypernetworks can be generated. For each synthetic hypernetwork $G$, we transform it into its 2-section network, and calculate the corresponding betweenness values $\mathbf{B}$ (where $\mathbf{B}_{v_i}$ denotes the betweenness value of node $v_i$). Then, multiple node pairs are sampled to construct ranking instances. For two random nodes $v_i$ and $v_j$, instance ($v_i$, $v_j$, $l_{v_iv_j}$) can be constructed as follows:
\begin{equation}
	\label{equ:samples label}
	l_{v_iv_j}=\left\{
	\begin{aligned}
		&1, \ \ \ \ \ \mathbf{B}_{v_i} > \mathbf{B}_{v_j} \\
		&0, \ \ \ \ \ \mathbf{B}_{v_i} < \mathbf{B}_{v_j}
	\end{aligned}
	\right.
\end{equation}
It is worth noting that ranking instances cannot be constructed when two nodes' betweenness values are equal. Therefore, this case should be avoid when sampling node pairs.

Through the method described above, each synthetic hypernetwork can generate one corresponding ranking sample set, and the training samples space $\Omega$ is also obtained:
\begin{equation}
	\label{equ:samples set}
	S_G=\{(v_0^G, v_1^G, l_{v_0v_1}^G), (v_1^G, v_2^G, l_{v_1v_2}^G), \dots\}
\end{equation}
\begin{equation}
	\label{equ:samples space}
	\Omega=\{S_{G_1}, S_{G_2}, \dots\}
\end{equation}

\subsection{Betweenness Approximation Model}
\label{sec:method-node approximation model}
To approximate the betweenness of a node in hypernetwork, we build a ranking model based on hypergraph neural network. For each ranking sample set $S_G$, our model first obtains the nodes' embeddings by a hypergraph neural network. Then, nodes' embeddings are feed into a fully-connected neural network to obtain the approximation betweenness values. The detailed process are introduced as follows.

For a given hypernetwork $G$, we first map the nodes into their dense vectors form with a hypergraph neural network. However, conventional transductive hypergraph neural networks such as HGNN and HyperGCN are not applicable in our framework due to the training process is conducted on a large number of small-scale synthetic hypernetworks, and then the trained model is applied real world hypernetworks. Thus, we choose the inductive hypergraph neural network, HyperSAGE \cite{hitter} which we proposed in our previous work. Generally, HyperSAGE contains two level of information aggregation, i.e., hyperedge level aggregation and node level aggregation.

In the step of hyperedge level aggregation, features of hyperedges are first obtained according to nodes' features as follows:
\begin{equation}
	\label{equ:node attention}
	\mathbf{A}^l={\rm softmax}(\mathbf{H}^T \odot (\mathbf{X}^l\mathbf{W}_1)^T)
\end{equation}
\begin{equation}
	\label{equ:hyperedge features}
	\mathbf{Y}^l=\mathbf{A}^l\mathbf{X}^l
\end{equation}
where $\mathbf{X}^l \in \mathbb{R}^{N \times D_l}$ and $\mathbf{Y}^l \in \mathbb{R}^{M \times D_l}$ are nodes' and hyperedges' embeddings in the $l$-th layer, respectively ($D_l$ denotes the embedding dimension in the $l$-th layer). $\mathbf{W}_1 \in \mathbb{R}^{D_l \times 1}$ is the trainable parameters. $\odot$ denotes the operation of element-wise product. Function $\rm softmax$ is used to normalize the aggregation weight. It is worth noting that due to the lack of initial node features $\mathbf{X}^0$, we set $\mathbf{X}^0=\mathbf{1}$. Based on the view of information propagation, nodes with high betweenness values as hubs of multiple shortest paths, have more powerful information propagation ability than those with low betweenness values. So, the initial nodes features $\mathbf{1}$ can be regarded as the initial information and each layer of hypergraph neural network is a kind of information propagation. After multiple layers, nodes' embeddings can reflect their ability of information propagation, and be further applied to betweenness approximation.

Once the hyperedges' embeddings are obtained, the hyperedge level aggregation can be performed as follows:
\begin{equation}
	\label{equ:hyperedge aggregation}
	\mathbf{Y}^{l+1}=f([\mathbf{Y}^l \mathbf{W}_2 || (\mathbf{H}^T \mathbf{H} \mathbf{Y}^l \mathbf{W}_3)] \mathbf{W}_4)
\end{equation}
where $\mathbf{W}_2$, $\mathbf{W}_3 \in \mathbb{R}^{D_l \times D_{l+1}}$ and $\mathbf{W}_4 \in \mathbb{R}^{2D_{l+1} \times D_{l+1}}$ are all trainable parameters. $||$ denotes the operation of matrix concatenation. Non-linear activation function $f$ is specified as ReLU.

The next step is node level aggregation. In this step, nodes aggregate information from their adjacent hyperedges as follow:
\begin{equation}
	\label{equ:node aggregation}
	\mathbf{X}^{l+1}=f([\mathbf{X}^l \mathbf{W}_5 || \mathbf{H} \mathbf{Y}^{l+1} \mathbf{W}_6] \mathbf{W}_7)
\end{equation}
where $\mathbf{W}_5 \in \mathbb{R}^{D_l \times D_{l+1}}$, $\mathbf{W}_6 \in \mathbb{R}^{D_{l+1} \times D_{l+1}}$ and $\mathbf{W}_4 \in \mathbb{R}^{2D_{l+1} \times D_{l+1}}$ are trainable parameters.

After multiple layers are chained in HyperSAGE, nodes' final embeddings $\mathbf{X}^L$ are obtained ($L$ denotes the total number of layers in HyperSAGE). Through feeding them into a fully-connected neural network, the approximated betweenness values of nodes can be calculated as follows:
\begin{equation}
	\label{equ:betweenness}
	\hat{\mathbf{B}}=f(\mathbf{X}^{L} \mathbf{W}_8 + b)
\end{equation}
where $\mathbf{W}_8 \in \mathbb{R}^{D_L \times 1}$ and $b \in \mathbb{R}$ are trainable parameters.

Through the above steps, the approximated betweenness values of nodes are obtained. According to the values, a hypernetwork can be dismantled by greedily removing nodes with the highest approximated betweenness values. After the removal of each batch of nodes, the approximated betweenness values of the residual nodes are re-calculated due to the structure changes of the hypernetwork. Betweenness value approximation and node removal are conducted repeatedly until the scale of GCC in hypernetwork decreased to a threshold value or all nodes are removed.

\subsection{Optimization}
\label{sec:method-optimization}
In order to optimize the betweenness approximation model, node ranking samples generated in Section \ref{sec:method-training samples generation} are utilized to update all trainable parameters in model. Thus, we build the pairwise ranking loss as follows.

For each samples set $S_G$, the Bayesian Personalized Ranking (BPR) \cite{bpr} loss of each instance $(v_i^G, v_j^G, l_{v_iv_j}^G)$ are calculated:
\begin{equation}
	\label{equ:predict label}
	\hat{l}^G_{v_iv_j}={\rm sigmoid}(\hat{\mathbf{B}}^G_{v_i} - \hat{\mathbf{B}}^G_{v_j})
\end{equation}
\begin{equation}
	\label{equ:simple bpr}
	loss_{v_i, v_j}^G = -l_{v_iv_j}^G {\rm log}(\hat{l}^G_{v_iv_j}) - (1 - l_{v_iv_j}^G) {\rm log}(\hat{l} - ^G_{v_iv_j})
\end{equation}
where $\hat{\mathbf{B}}^G_{v_i}$ denotes the approximated betweenness value of node $v_i$ in hypernetwork $G$. According to the pairwise loss of a single instance, the total loss of the whole samples $\Omega$ can be obtained as follows:
\begin{equation}
	\label{equ:total loss}
	loss = \frac 1 {|\Omega|} \sum_{S_G \in \Omega} \frac 1 {|S_G|} \sum_{(v_i^G, v_j^G, l_{v_iv_j}^G) \in S_G} loss^G_{v_i, v_j}
\end{equation}

\begin{algorithm}
	\caption{Training process of HND}
	\label{alg:training}
	\begin{algorithmic}[1]
		\REQUIRE Scale range of synthetic hypernetworks  $(N_{min}, N_{max})$, burning propabality range$(p_{min}, p_{max})$ in HyperFF, expanding propabality range $(q_{min}, q_{max})$ in HyperFF, synthetic hypernetwork number $J$, max iteration number $I$, HyperSAGE layer number $L$, embedding dimension $D$, node pairs sample ratio $r$
		\ENSURE Betweenness approximation model parameters $\mathbf{\Theta}$
		\STATE Initialize model parameters $\mathbf{\Theta}$ randomly
		\STATE Initialize sample set $\Omega = \{\}$
		\FOR{$n$ = 1 to $J$}
		\STATE Take random values $p, q, N$ in range $[p_{min}, p_{max}], [q_{min}, q_{max}], [N_{min}, N_{max}]$, respectively
		\STATE Generate a hypernetwork $G$ which has $N$ nodes by HyperFF with hyperparameters $p$ and $q$
		\STATE Calculate the exact betweenness values $\mathbf {\hat{B}}^G$ of $G$'s corresponding 2-section network
		\STATE Sample $\lceil rN \rceil$ node pairs randomly
		\STATE Generate ranking samples $S_G$ according to Equations (\ref{equ:samples label}) and (\ref{equ:samples set})
		\STATE Add $S_G$ into sample set $\Omega$
		\ENDFOR
		\FOR{$i$ = 1 to $I$}
		\FOR{$S_G \in \Omega$}
		\STATE Embed nodes in hypernetwork $G$ according to Equations (\ref{equ:node attention}) - (\ref{equ:node aggregation})
		\STATE Calculate nodes' approximated betweenness values according to Equation (\ref{equ:betweenness})
		\STATE Calculate pairwise loss according to Eqaution (\ref{equ:simple bpr})
		\STATE Update model parameters $\mathbf{\Theta}$ through gradient decrease
		\ENDFOR
		\ENDFOR
		\RETURN Model parameters $\mathbf{\Theta}$
	\end{algorithmic}
\end{algorithm}

Then gradient decrease \cite{adam} is utilized to update all trainable parameters in our model. Through multiple iterations of parameters updating, the loss of model will converge. Then, the model can be used to approximate nodes' betweenness in real world hypernetworks.

\subsection{Time Complexity}
In the process of inference, the time complexity of HND is mainly from hypernetwork embedding (i.e., HyperSAGE) and the approximation function (i.e., the fully-connected neural network). For the former part, there are two steps in each layer of HyperSAGE: (1) In the step of hyperedge level aggregation, the time complexity of attention mechanism and information propagation are both $O(MD)$; (2) In the step of node level aggregation, the time complexity of information propatgation is $O(ND)$. Therefore, the time complexity of HyperSAGE is $O((N+2M)LD)$. For the latter part, the time complexity of fully-connected layer is $O(ND)$. So, the inference complexity of HND is $O((N+2M)LD+ND)$. While for exact betweenness centrality calculation, its time complexity for unweighted network is $O(NM)$. Because $L, D \ll N, M$ in most real world scenarios, comparing with exact betweenness centrality calculation, our proposed HND has a linear time complexity and is more suitable for application.

\section{Experiments}
\label{sec:experiments}
In this section, we conduct extensive experiments to verify the effectiveness of our proposed method.

\subsection{Experimental Datasets and Settings}
\subsubsection{Experimental Datasets}
Five real world hypernetworks are collected to evaluate our proposed method. Brief introductions about these datasts are listed below.
\begin{itemize}
	\item \textbf{Cora-co-authorship} \cite{HyperGCN} This dataset contains scientific papers published in the field of machine learning. We construct hypernetwork by taking authors as nodes and co-author relations as hyperedges.
	\item \textbf{Citeseer} \cite{HyperGCN} This dataset contains scientific papers in six fields and their citation relations. To construct hypernetwork, papers and citation relations are mapped as nodes and hyperedges, respectively.
	\item \textbf{MAG} \cite{data2} This dataset contains scientific papers and authors from the field of history in Microsoft Academic Graph. Similar to Cora-co-authorship, papers are considered as nodes and co-author relations are mapped as hyperedges.
	\item \textbf{Pubmed} \cite{HyperGCN} This dataset contains scientific papers from the field of diabetes and their citation relations. Due to the form of Pubmed is the same as Citeseer, the way of constructing hypernetworks in Citeseer is also suitable for Pubmed.
	\item \textbf{NDC} \cite{data1} This dataset contains many drugs and each drug is consisted of multiple substances. Through considering substances and drugs as nodes and hyperedges, respectively, hypernetwork can be constructed.
\end{itemize}
\begin{table*}[htbp!]
	\setlength{\belowcaptionskip}{0.2cm} 
	\caption{The statistics of datasets}
	\centering
	\renewcommand\arraystretch{1.0}
	\begin{tabular}{cccccc}
		\toprule
		Datasets                 & Cora-co-authorship & Citeseer & MAG  & NDC   & Pubmed \\ \toprule
		Number of Nodes          & 1676               & 1019     & 1669 & 3065  & 3825   \\
		Number of Hyperedges     & 463                & 626      & 784  & 4533  & 5432   \\
		Average Node Hyperdegree & 1.66               & 2.23     & 1.59 & 13.57 & 7.45   \\
		Average Hyperedge Size   & 6.00               & 3.63     & 3.38 & 9.17  & 5.25   \\ \toprule
	\end{tabular}
	\label{tab:dataset}
\end{table*}

For these datasets, we conducted some pre-process steps. Due to the dismantling problem concentrates on the connectivity of network, so a disconnected network will disturb experimental results. Therefore, for each original hypernetwork, we only select their GCCs as the initial hypernetwork waiting to be dismantled. Some statistics of hypernetworks after pre-process are summarized in Table \ref{tab:dataset}.

\subsubsection{Baselines}
To evaluate the dismantling performance of our method, various baselines are selected and their introduction are listed below:
\begin{itemize}
	\item \textbf{Highest Degree Adaptive (HDA)} This method dismantles a network according to degree. In each removal step, nodes with the highest degree will be removed. After each removal step, nodes' degree will be re-calculated due to the changes of network structure.
	\item \textbf{Highest Hyperdegree Adaptive (HHDA)} This method dismantles hypernetwork according to hyperdegree. Similar to HDA, nodes with the highest hyperdegree will be removed in each step, and nodes' hyperdegree will be updated after removal.
	\item \textbf{Collective Influence (CI)} \cite{CI} This method removes nodes according to their CI values, and the way of calculating CI value for each node is:
	\begin{equation}
		\label{equ:ci}
		CI_{v_i} = (deg(v_i) - 1)\sum_{v_j \in Nei_k(v_i)}(deg(v_j) - 1)
	\end{equation}
	where $Nei_k(v_i)$ denotes the $k$-hop neighbors of node $v_i$ and in this paper we set the $k$ is 2. Node with the highest CI value will be removed in each step, and the CI values of residual nodes will also be re-calculated.
	\item \textbf{GND} \cite{gnd} GND first computes the network's weighted laplacian matrix which is utilized to obtain nodes' eigenvector through spectrum approximation. Then, nodes are splited into two groups according to their eigenvector. Finally, weighted vertex cover algorithm is used to select nodes to be removed.
	\item \textbf{FINDER} \cite{finder} This method is based on deep reinforcement learning. It builds an agent and makes it do dismantling exercises on lots of small scale synthetic networks to optimize the agent's dismantling strategy. Once its strategy converges, the agent can be used to dismantle real world networks.
	\item \textbf{SubTSSH} \cite{SubTSSH} This method is designed for the problem of selecting target nodes set in hypernetworks. Through iteratively conduct node remeval and influence propagation, it can be used for hypernetwork dismantling.
	\item \textbf{HITTER} \cite{hitter} This method is specifically designed for hypernetwork dismantling based on deep reinforcement learning. It first builds an agent to do trail-and-errors on many small scale synthetic hypernetworks. After the agent's dismantling strategy is well optimized, it can be utilized to dismantle real world hypernetworks.
\end{itemize}

In these above baselines, HDA, HHDA, CI and SubTSSH are methods based on centrality measure, GND is the recently proposed heuristic methods, while FINDER and HITTER are methods based on deep learning. Generally, these baselines cover several commonly used dismantling methods. Moreover, some of these baselines (HDA, CI, GND and FINDER) cannot be directly applied to hypernetworks. Therefore, we transform hypernetworks into their corresponding 2-section network when applying these methods.

\subsubsection{Metrics}
Accumulated normalized connectivity (ANC) \cite{ANC} is adopted to evalute performance of hypernetwork dismantling. The calculation of ANC is:
\begin{equation}
	\label{equ:anc}
	ANC(\kappa) = \frac 1 K \sum_{k=1}^K \frac {connectivity(G \backslash \{v_1, v_2, \dots, v_k\})} {connectivity(G)}
\end{equation}
where $\kappa = \{v_1, v_2, \dots, v_K\}$ is the nodes removal sequence obtained by various dismantling methods. $G \backslash \{v_1, v_2, \dots, v_k\}$ denotes the residual hypernetwork after removing nodes $\{v_1, v_2, \dots, v_k\}$ from $G$. Given a nodes removal sequence $\kappa$, a low ANC value means this sequence can dismantle network effectively.

\subsubsection{Experimental Settings}
In the step of sample generation, the minimal scale of synthetic hypernetwork $N_{min}$ is set to 100 and the maximal scale $N_{max}$ is set to 150. The burning probability range $(p_{min}, p_{max})$ and expanding probability range $(q_{min}, q_{max})$ are all fixed to $(0.1, 0.4)$. The number of synthetic hypernetworks $J$ is set to 1000, and for each hypernetwork we randomly sample $\lceil 0.9N \rceil$ node pairs as training samples. In the betweenness approximation model, the layer of HyperSAGE is set to 4 and the embedding dimension is set to 32. In training process, the learning rate of model is set to 0.005. And 50 synthetic hypernetworks are also generated as validation dataset to verify the training of model. Early stopping mechanism is adopted to avoid the over-fitting of model, and its patience is fixed to 100.

\subsection{Experimental Results and Analyses}
\subsubsection{Overall Performance}
\begin{table*}[htbp!]
	\setlength{\belowcaptionskip}{0.2cm} 
	\caption{The overall performance}
	\centering
	\renewcommand\arraystretch{1.5}
	\begin{tabular}{lllllllll}
		\toprule
		Datasets           & HDA    & CI     & GND    & FINDER & HHDA   & SubTSSH & HITTER & HND    \\ \toprule
		Cora-co-authorship & 0.1564 & 0.1181 & 0.1111 & 0.4068 & 0.0977 & 0.1267  & 0.0792 & \textbf{0.0713} \\
		Citeseer           & 0.0930 & 0.0915 & 0.2528 & 0.1109 & 0.0788 & 0.0965  & 0.0607 & \textbf{0.0592} \\
		MAG                & 0.0261 & 0.0238 & 0.0335 & 0.0410 & 0.0195 & 0.0226  & \textbf{0.0130} & 0.0161 \\
		Pubmed             & 0.3933 & 0.3930 & 0.3606 & 0.4809 & 0.3831 & 0.4038  & 0.3529 & \textbf{0.3344} \\
		NDC                & 0.2608 & 0.2623 & 0.4372 & 0.4804 & 0.2374 & 0.2666  & 0.2209 & \textbf{0.2144} \\ \toprule
	\end{tabular}
	\label{tab:results}
\end{table*}
\begin{figure*}[htbp!]
	\centering
	\subfigtopskip=0pt
	\subfigure[{Cora-co-authorship}]{
		\includegraphics[scale=0.2]{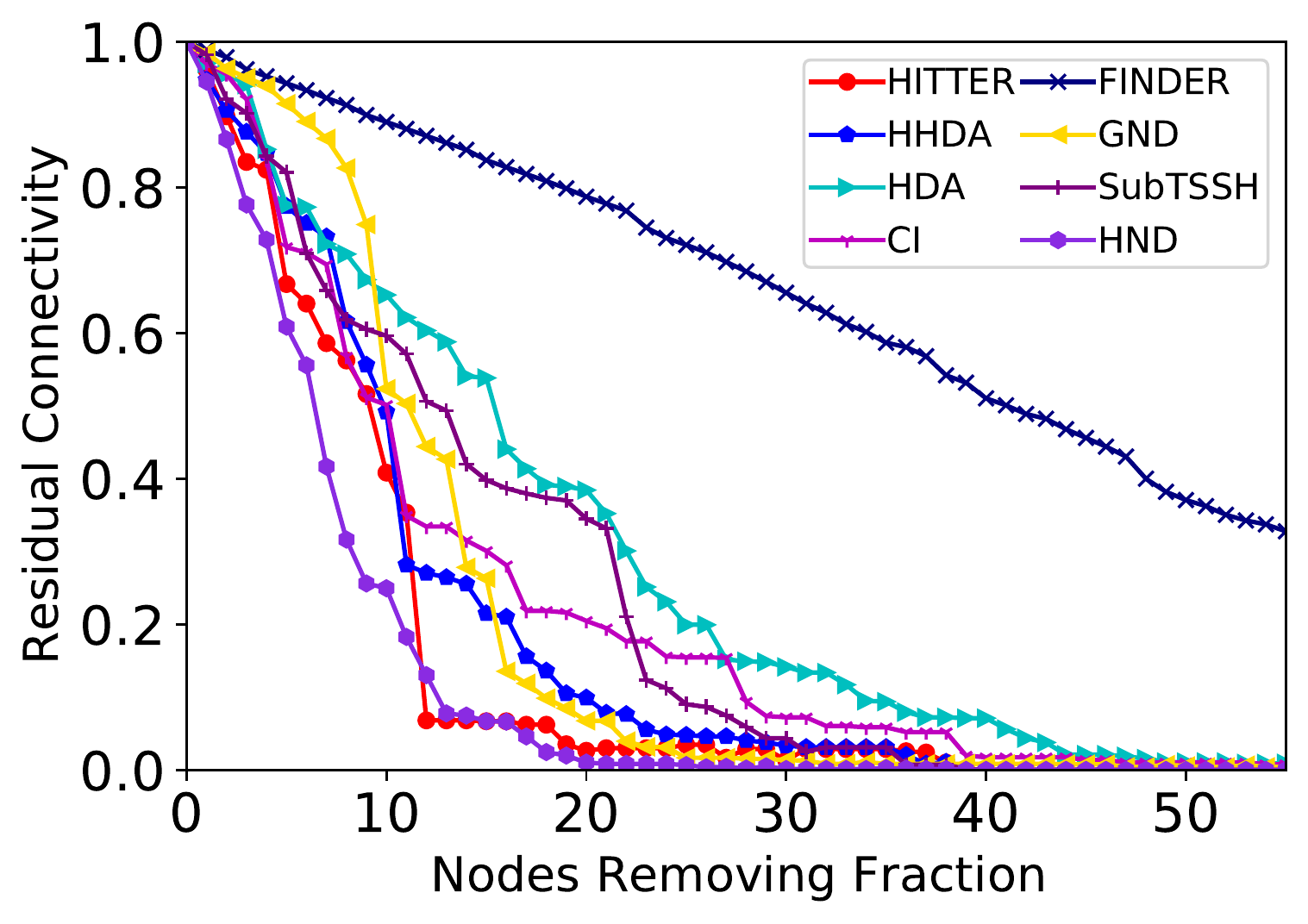}
		\label{subfig:cora-hnd}
	}
	\subfigure[MAG]{
		\includegraphics[scale=0.2]{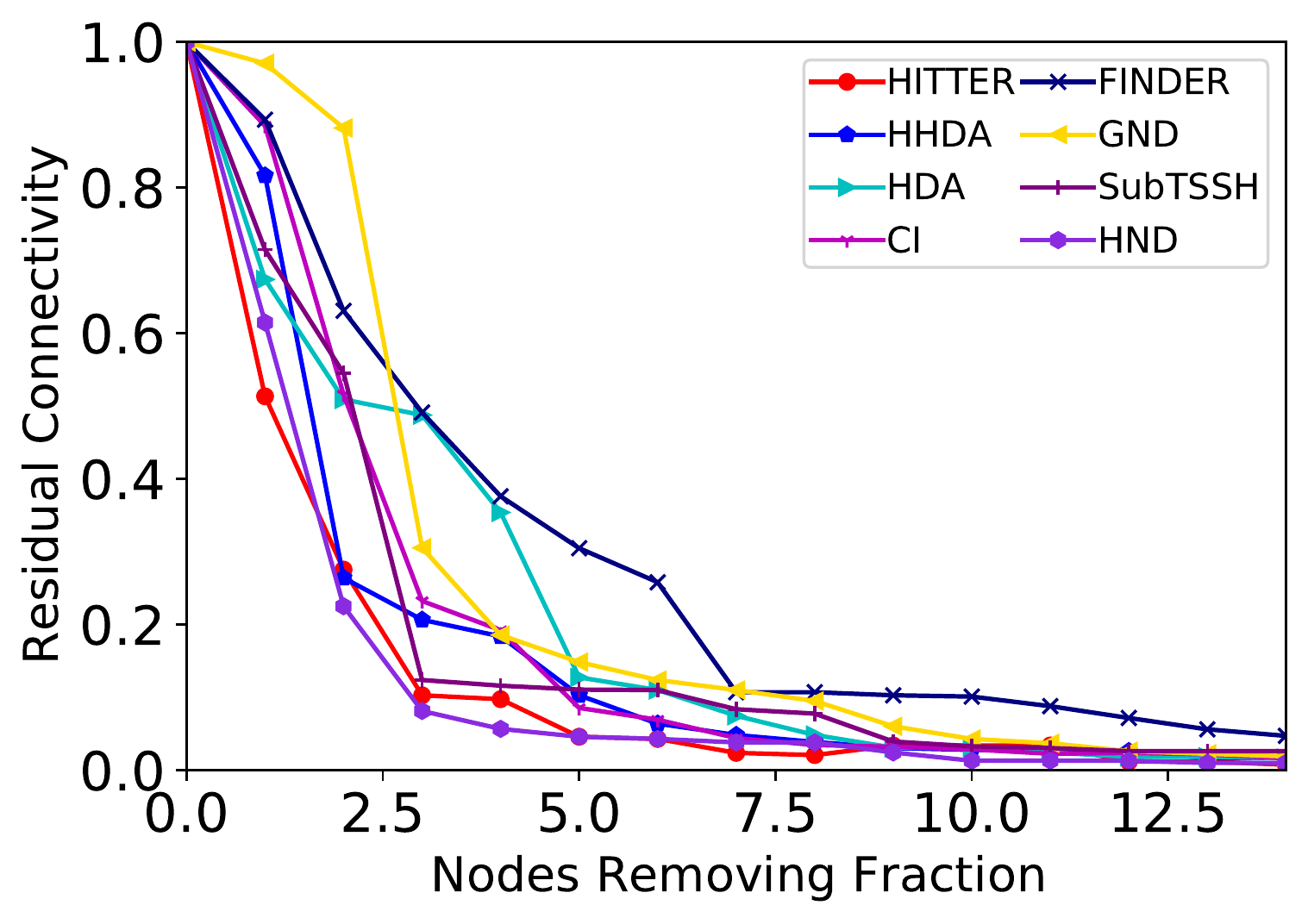}
		\label{subfig:mag-hnd}
	}
	\subfigure[Citeseer]{
		\includegraphics[scale=0.2]{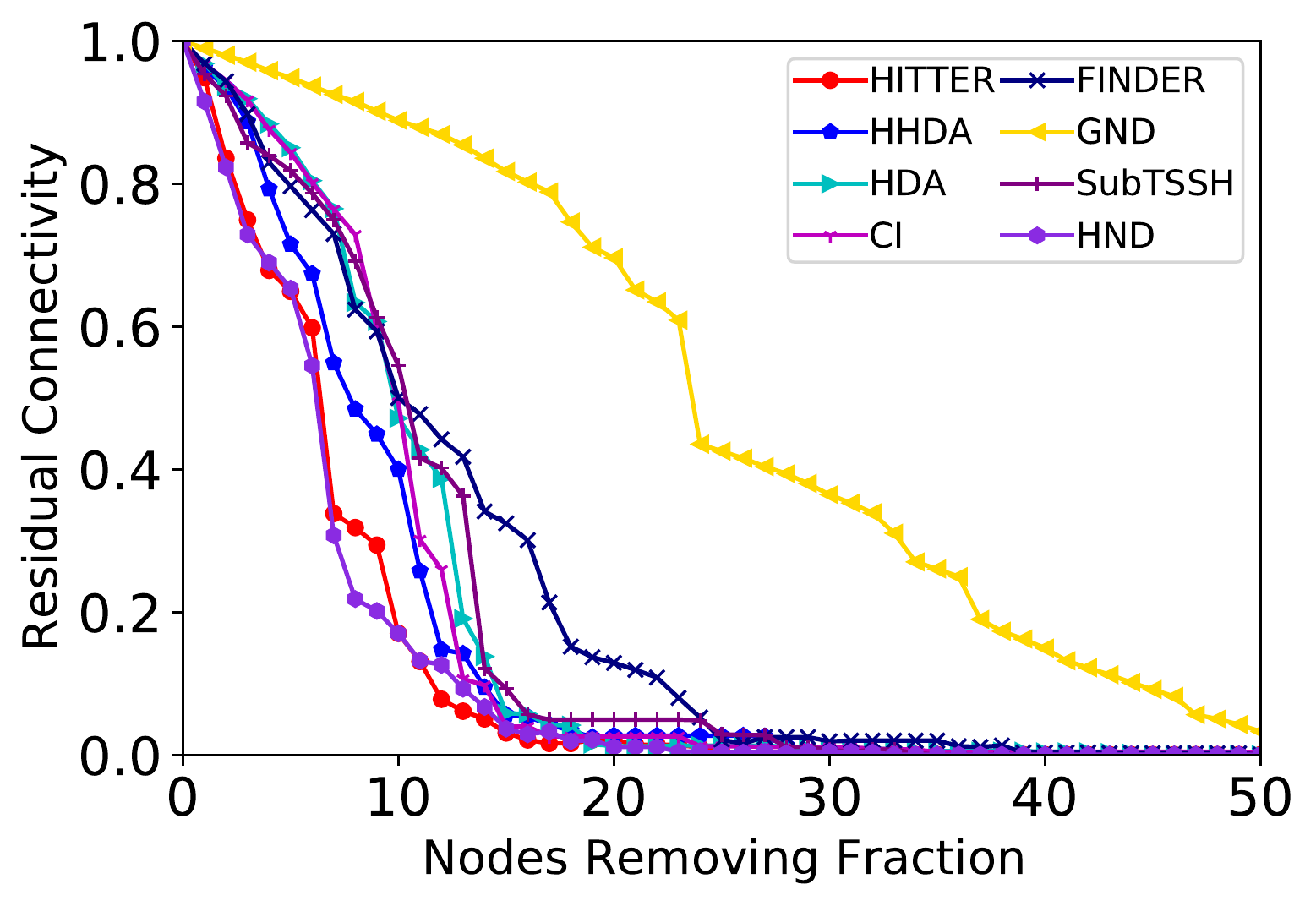}
		\label{subfig:citeseer-hnd}
	}
	\subfigure[NDC]{
		\includegraphics[scale=0.2]{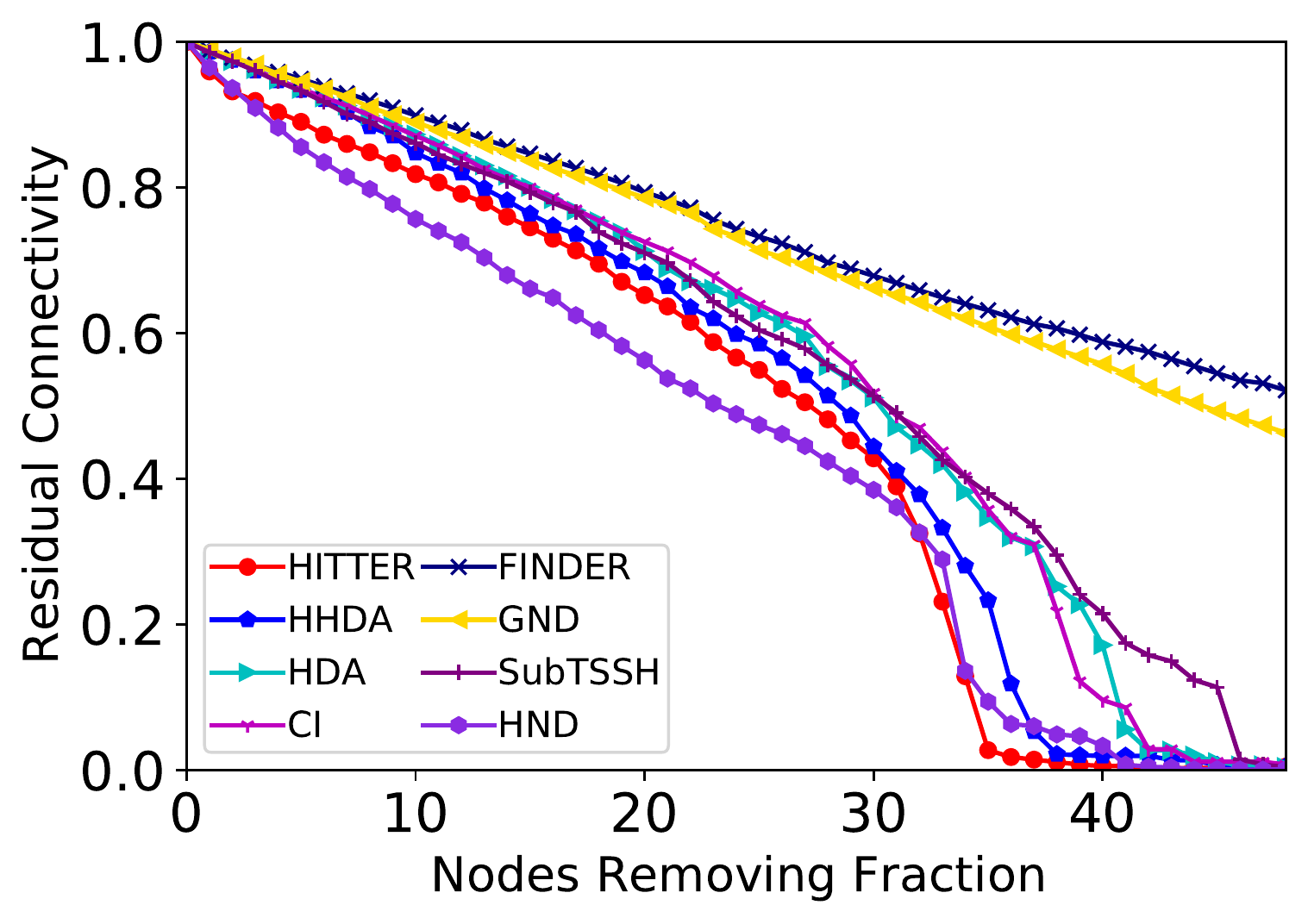}
		\label{subfig:ndc-hnd}
	}
	\subfigure[Pubmed]{
		\includegraphics[scale=0.2]{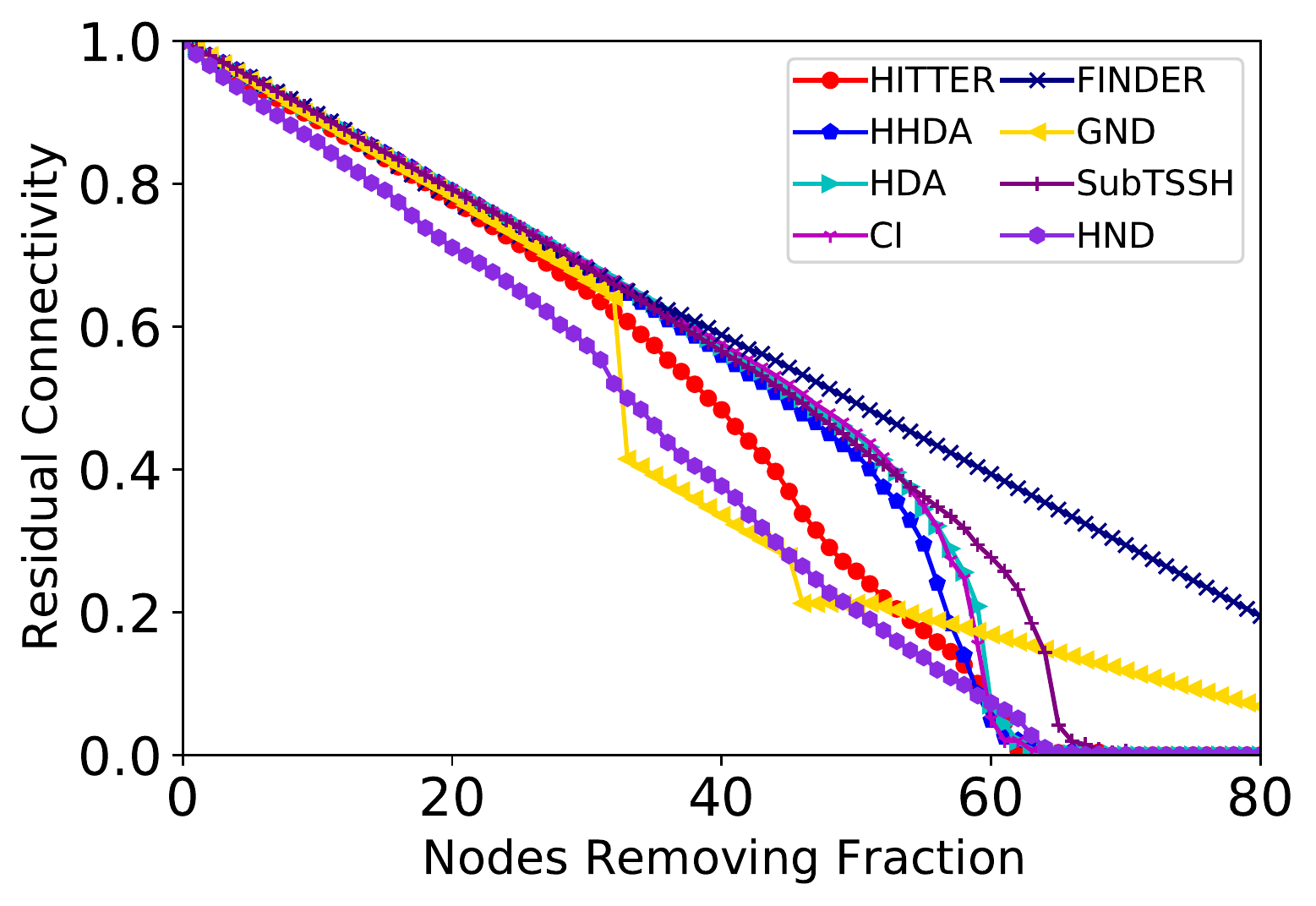}
		\label{subfig:pubmed-hnd}
	}
	\caption{Detailed dismantling curve}
	\label{fig:anc}
\end{figure*}
The proposed HND and baselines are applied to dismantle those real world hypernetworks in datasets, and their performance are summarized in Table \ref{tab:results}. According to the results, we can easily conclude that our proposed HND has a significant dismantling performance comparing with baselines in the majority of datasets. This indicates the effectiveness of approximating betweenness for hypernetwork dismantling with an advantage of reducing computational complexity. We further show the detail of dismantling process in Fig. \ref{fig:anc}. The detailed ANC curves in Fig. \ref{fig:anc} show for most hypernetworks the removal of only a small portion of nodes will significantly destroy their connectivity, and under a same removal budget (i.e., removing fraction) our proposed HND often fragment the hypernetwork into the smallest connected components.

It is also worth noting that the performance of dismantling methods designed for traditional simple network, i.e., HDA, CI, GND and FINDER, is generally poorer than those designed specifically for hypernetwork, i.e., HHDA, SubTSSH and HITTER. This is because hypernetwork must be transformed to its 2-section form before applying dismantling methods designed for traditional simple network, and this kind of transformation introduces some noisy structure like dense clique. Moreover, FINDER as the SOTA method for traditional network dismantling to our best of knowledge, performs even worse than the other traditional simple network dismantling methods. The reason maybe that it utilizes BA model \cite{ba} for synthetic network generation in the training process. However, BA model is a generative model for traditional simple network and it cannot model the structure of hypernetwork.

\subsubsection{Analysis of Model Efficiency}
The motivation of proposing HND is to reduce high computational complexity of global structure-based centrality like betweenness while preserving its effectiveness in hypernetwork dismantling. 

\begin{figure}[htbp!]
	\centering
	\includegraphics[scale=0.3]{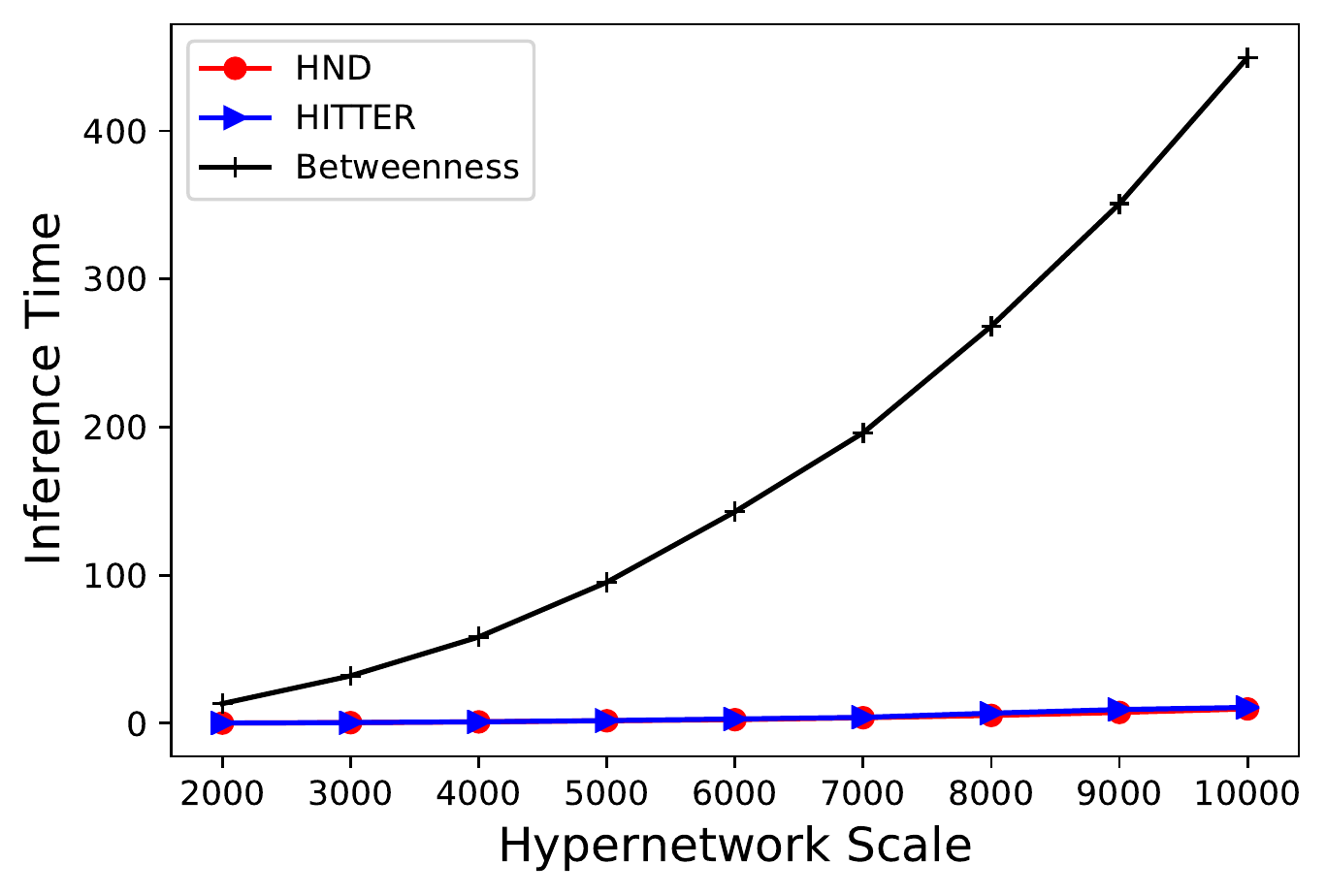}
	\caption{Inference time under different hypernetwork scales}
	\label{fig:infrence efficiency}
\end{figure}

First, the inference efficiency of HND is compared against HITTER and the exact betweenness centrality under different hypernetwork scales and the time consumption of each calculation in these methods is shown in Fig. \ref{fig:infrence efficiency}. From the results shown in Fig. \ref{fig:infrence efficiency}, we can find that the inference time of HND and HITTER increases linearly with the hypernetwork scale while the inference time of the exact betweenness centrality grows quadratically. This indicates a potential for applying our proposed HND to large scale hypernetwork.

\begin{table}
	\setlength{\belowcaptionskip}{0.2cm} 
	\caption{Efficiency comparison between HITTER and HND}
	\centering
	\renewcommand\arraystretch{1.0}
	\begin{tabular}{ccc}
		\toprule
		Method           & Training Iterations  & Training Time (s) \\ \toprule
		HITTER & 84766($\pm$17470) & 32505.51($\pm$2808.82) \\
		HND           & 101($\pm$45) & 11235.07($\pm$4866.57) \\ \toprule
	\end{tabular}
	\label{tab:training efficiency}
\end{table}

Moreover, as HND and HITTER are deep learning-based hypernetwork dismantling methods, their training efficiency is also compared. From the results shown in Table \ref{tab:training efficiency}, we can see that the training of HITTER usually needs tens of thousands of iterations while HND only needs hundreds of iterations. The training time of HND is also obviously less than HITTER. The core reason behind this is the difference of training mode. Reinforcement learning adopted by HITTER trains the model with exploration and exploitation and huge time is needed to explore effective dismantling actions while supervised learning adopted by our HND directly guides the model training with effective dismantling actions. Thus, HND can significantly decrease the training time comparing with HITTER.

\subsubsection{Impact of the number of embedding layers}
\begin{figure}
	\centering
	\subfigtopskip=0pt
	\subfigure[{Cora-co-authorship}]{
		\includegraphics[scale=0.26]{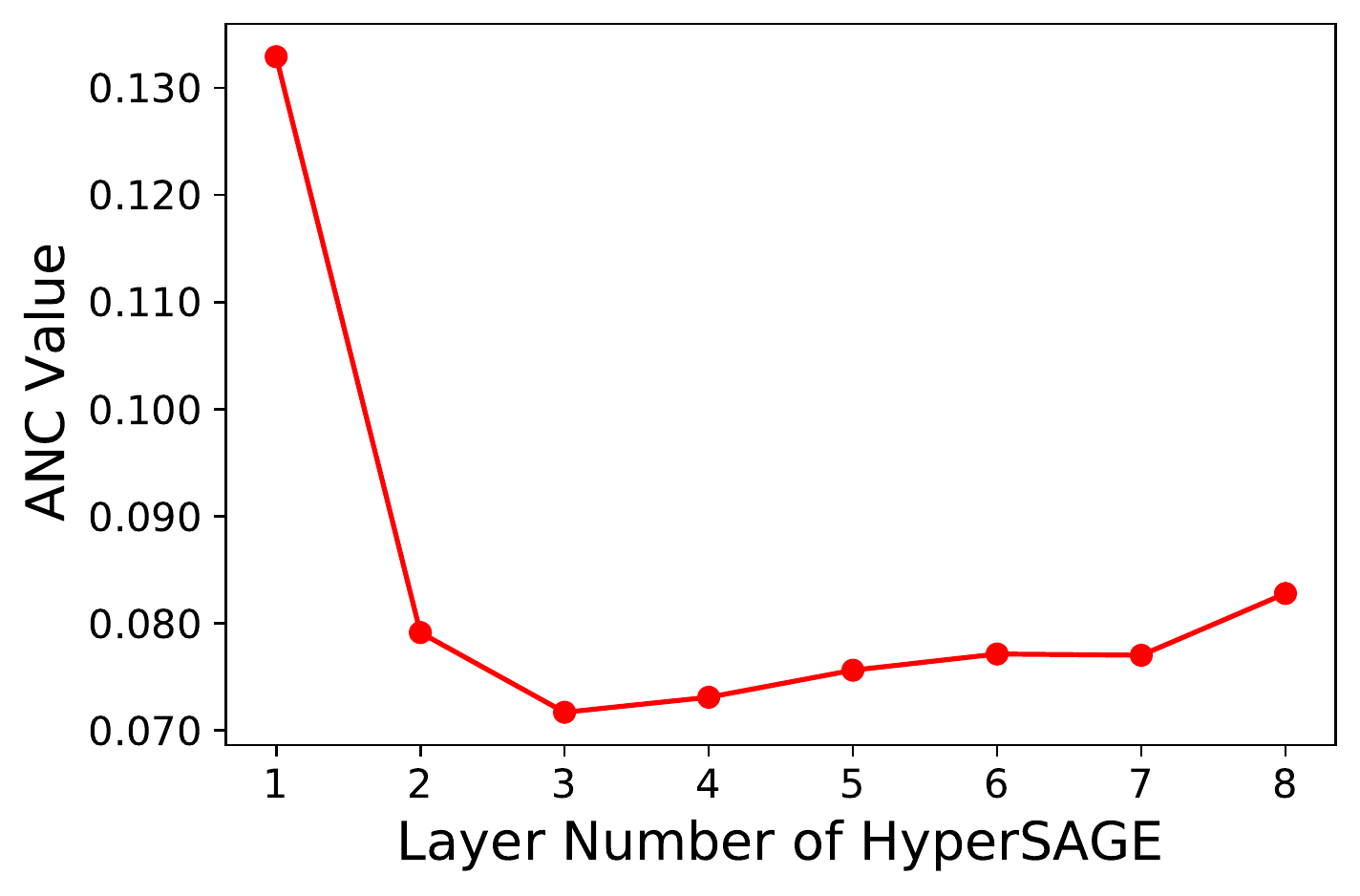}
		\label{subfig:cora-layer}
	}
	\subfigure[MAG]{
		\includegraphics[scale=0.26]{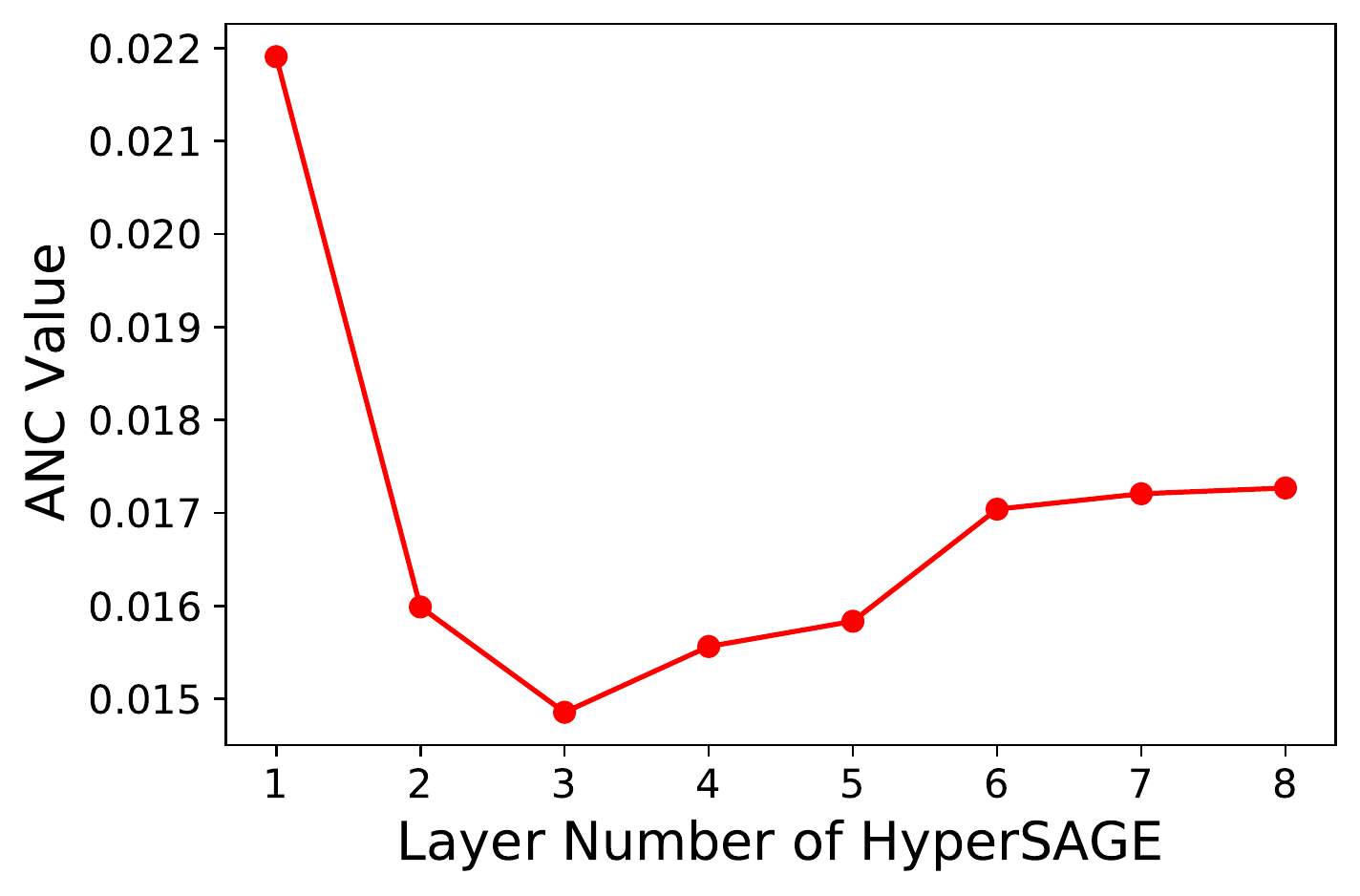}
		\label{subfig:mag-layer}
	}
	\subfigure[Citeseer]{
		\includegraphics[scale=0.26]{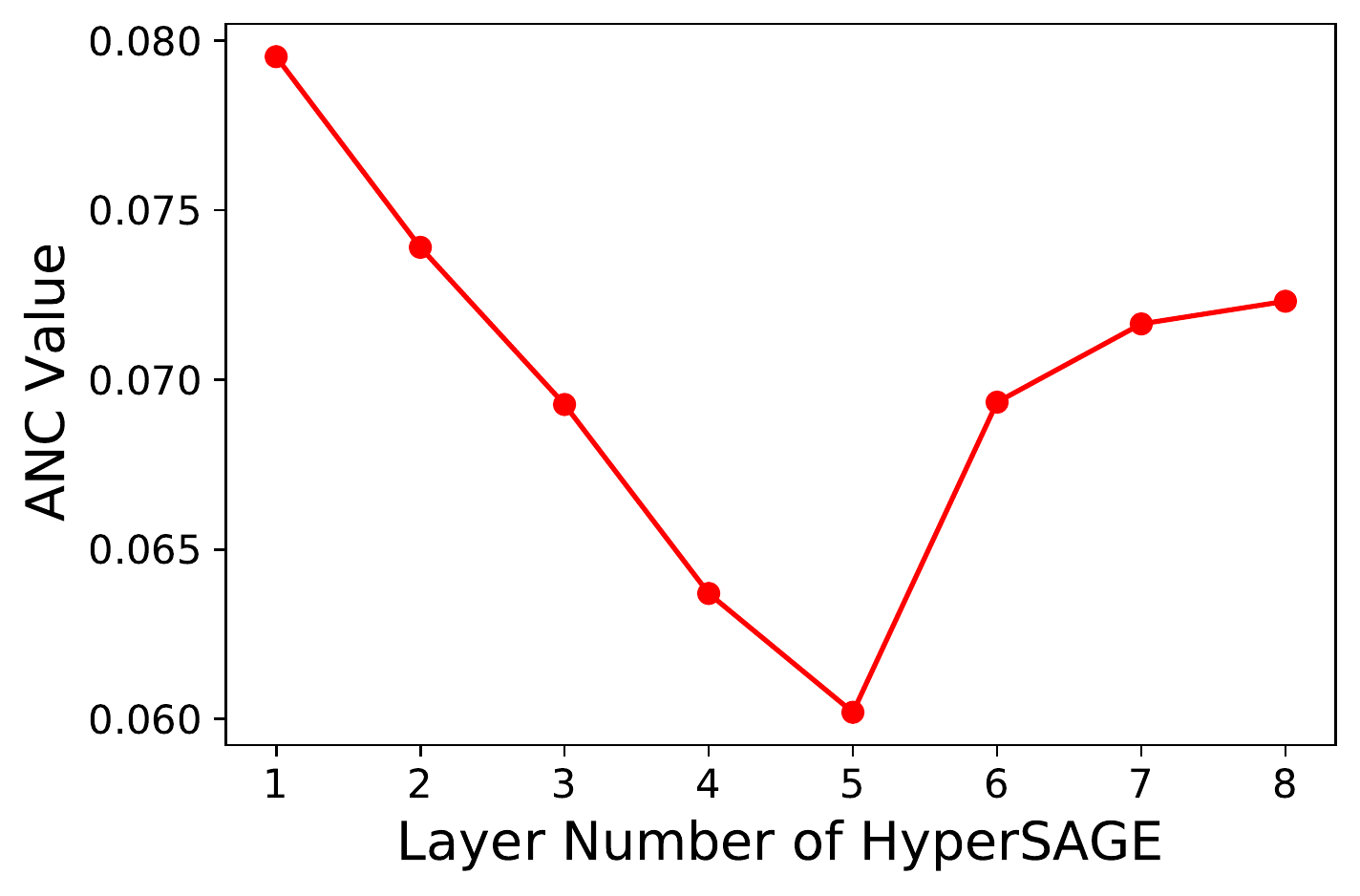}
		\label{subfig:citeseer-layer}
	}
	\subfigure[NDC]{
		\includegraphics[scale=0.26]{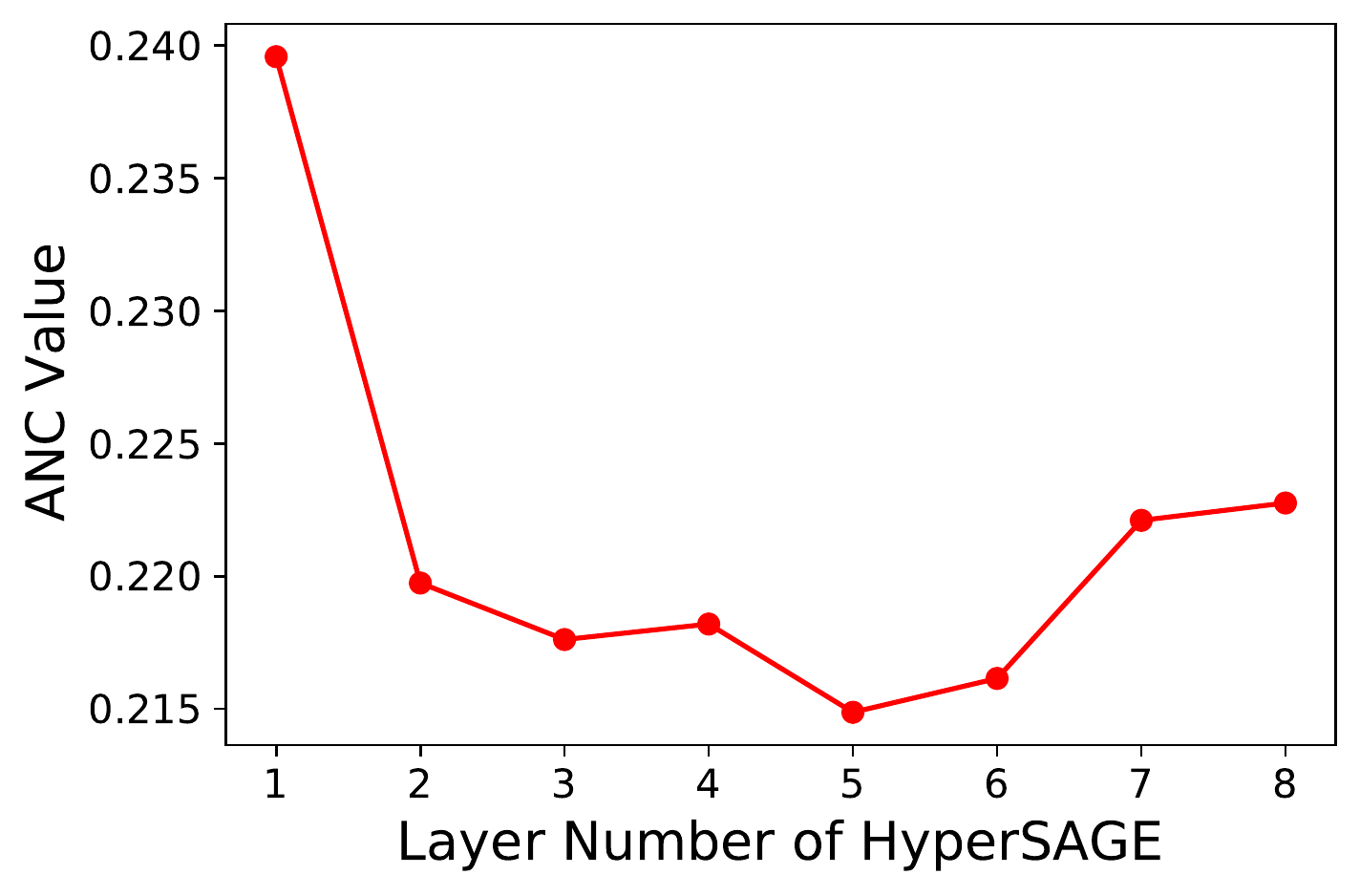}
		\label{subfig:ndc-layer}
	}
	\subfigure[Pubmed]{
		\includegraphics[scale=0.26]{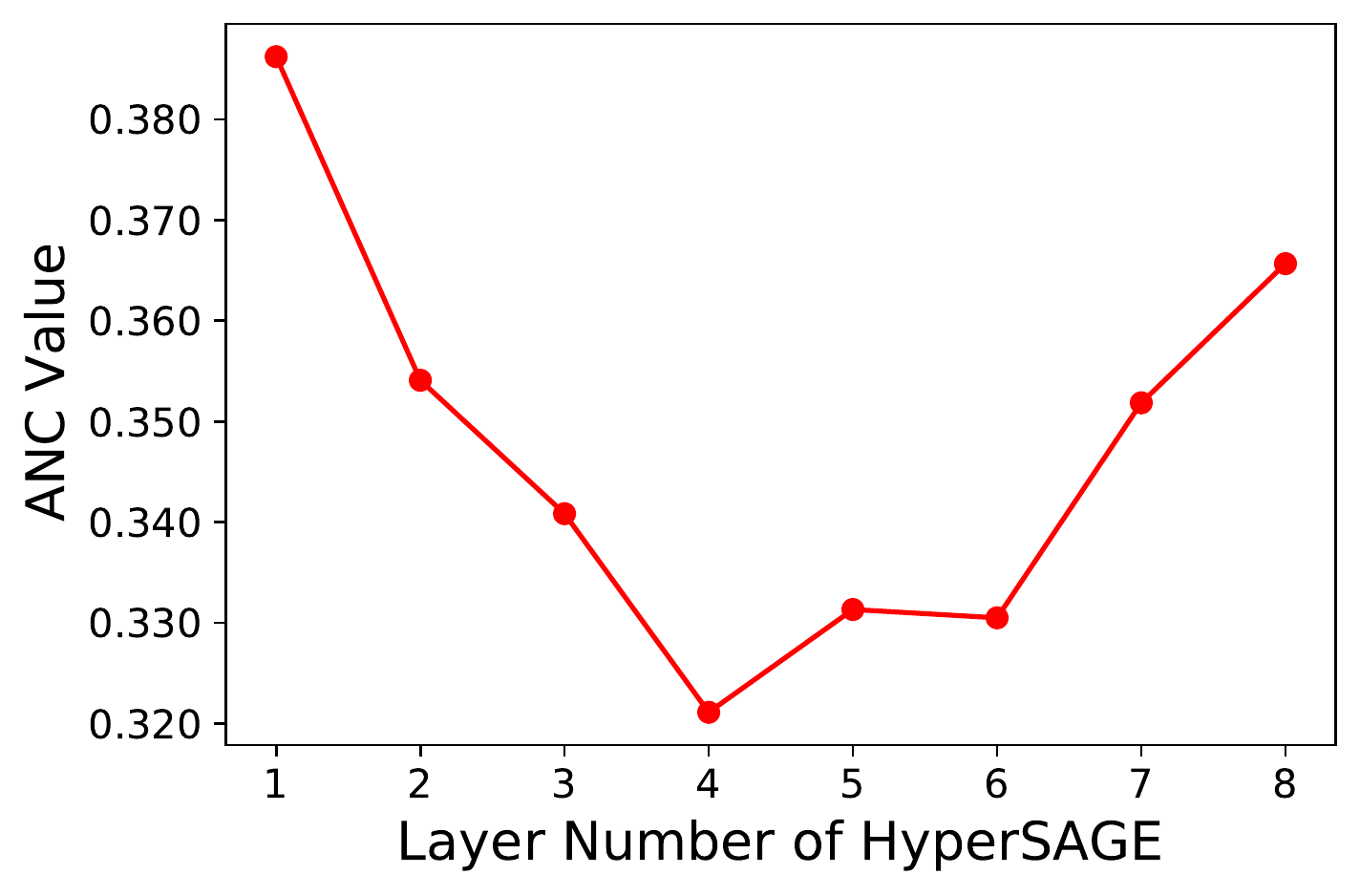}
		\label{subfig:pubmed-layer}
	}
	\caption{Impact of the number of embedding layers}
	\label{fig:layer number}
\end{figure}

HND relies on a multiple layer hypergraph neural network to extract structural information into node embeddings which are then utilized for betweenness approximation. Different number of layers indicates different awareness ability of global hypernetwork structure, thus the number of layers $L$ has a vital influence on hypernetwork dismantling performance. The impact of the number of embedding layers on each dataset is shown in Fig. \ref{fig:layer number}. From this figure, we can conclude that a too small or too big $L$ value will reduce the hypernetwork dismantling performance. The reason may be too few layers will limit the information propagation of hypergraph neural network, which prevents the node embeddings from preserving enough global structure information while too many layers will over smooth the node embeddings and make nodes less distinguishable. In addition, too many layers also bring extra computational cost. Thus, taking both the effectiveness and efficiency into account, it is necessary to choose a proper number of embedding layers in real application.

\section{Conclusion}
\label{sec:conclusion}
\noindent
In this paper, we proposed betweenness approximation-based hypernetwork dismantling method, named \emph{HND}. For achieving betweenness approximation in a supervised manner, we utilized a hypergraph neural network to preserve structure information into the node embeddings. In order to achieve the real world hypernetwork dismantling, we trained the model based on a large number of synthetic hypernetworks with supervision of exact betweenness value. Extensive experiments conducted on the five real world hypernetworks demonstrated that our proposed HND outperforms baselines in terms of the effectiveness and the time efficiency. 

In future work, we will consider the situation of noisy hyperedges in hypernetwork and design robust leaning-based hypernetwork dismantling methods.

\vskip 2mm
\large
\noindent
\textbf{Acknowledgment}
\vskip 2mm

\Acknow
\noindent
This work is partially supported by the National Natural Science Foundation of China under Grant (62006003), the Natural Science Foundation of Anhui Province (2208085QF197), the Hefei Key Common Technology Project under Grant (GJ2022GX15). The author is grateful to the Institute of Data Intelligence and  Social Computing in Anhui University for funding this research.

\vskip 2mm
\normalsize
\noindent
\renewcommand\refname{\large

\begin{biography}[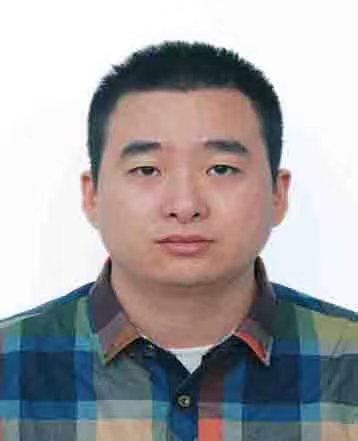]
\noindent
\textbf{Yaguang Guo} is a associate professor at the School of Management at the Hefei University of Technology of China.
\end{biography}
\vskip 18mm
\begin{biography}[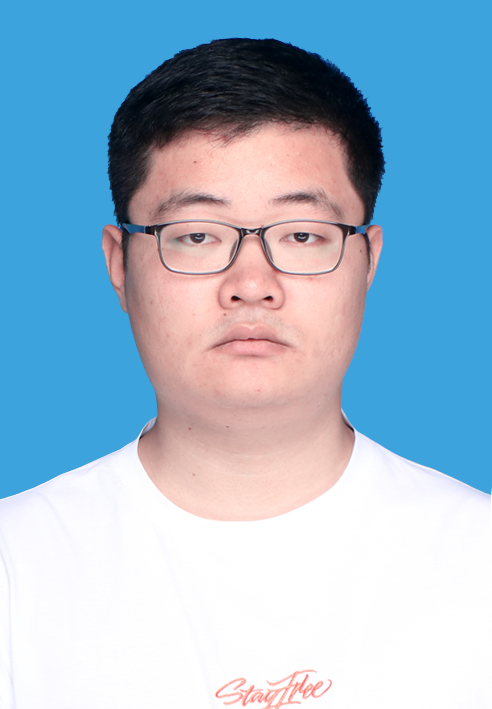]
\noindent
\textbf{Wenxin Xie} received the B.S. degree from Liaoning Petrochemical University, Fushun, China, in 2018. He is currently pursuing the M.S. degree in computer science and technology with Anhui University, Hefei, China. His research interest includes graph combinatorial optimization.
\end{biography}
\begin{biography}[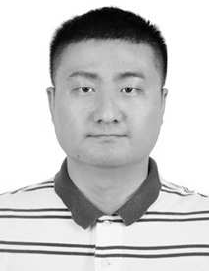]
\noindent
\textbf{Qingren Wang} is a lecturer in the School of Computer Science and Technology at Anhui University, Hefei, China. His current research interests include personalized recommendation, service computing, edge computing, crowdsourcing and data mining.
\end{biography}

\begin{biography}[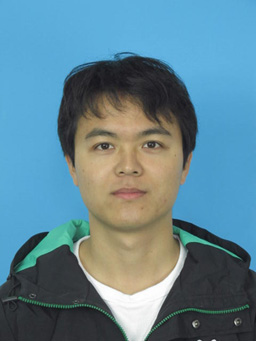]
\noindent
\textbf{Dengcheng Yan} received the B.S. and Ph.D. degrees from the University of Science and Technology of China, Hefei, China, in 2011 and 2017, respectively. From 2017 to 2018, he was a Core Technology Researcher and a Big Data Engineer with the Research Institute of Big Data, iFlytek Company Ltd, Hefei. He is currently a Lecturer with Anhui University. His research interests include software engineering, recommendation systems, and complex networks.
\end{biography}

\begin{biography}[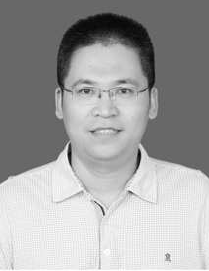]
\noindent
\textbf{Yiwen Zhang} is a professor in the School of Computer Science and Technology at Anhui University. He received his PhD degree from the School of Management, Hefei University of Technology, Hefei, China. His current research interests include personalized recommendation, service computing, edge computing and data mining.
\end{biography}

\begin{strip}
\end{strip}

  \end{document}